\documentstyle[prb,preprint,aps]{revtex}

\tighten
\begin{document}
\title{{\bf {Dynamic instability in resonant tunneling}}}
\author{G. Obermair$^{1}$, J. Inkoferer$^{1}$ and F. Claro$^{2}$}
\address{$^{1}${\it Fakultat Physik, Universit\"at Regensburg, }\\
D93040 Regensburg, Germany\\
$^{2}$ Facultad de F\'{\i}sica, Pontificia Universidad Cat\'{o}lica de\\
Chile,Casilla 306, Santiago 22, Chile}
\address{}
\address{\parbox{14cm}{\rm PACS numbers:  73.40.Gk,73.40.-c,72.15.Gd}}
\maketitle

\begin{abstract}
We show that an instability may be present in resonant tunneling through a
quantum well in one, two and three dimensions, when the resonance lies near
the emitter Fermi level. A simple semiclassical model which simulates the
resonance and the projected density of states by a nonlinear conductor, the
Coulomb barrier by a capacitance, and the time evolution by an iterated map,
is used. The model reproduces the observed hysteresis in such devices, and
exhibits a series of bifurcations leading to fast chaotic current
fluctuations.
\end{abstract}

\newpage

Resonant tunneling through a double barrier, originally a rather elementary
academic exercise, has already for over three decades shown to be an
extremely rich source of new and surprising physics.\cite{tsu} One instance
is the bistable region in the I-V characteristic that some samples exhibit,
leading to hysteresis as the curve is traced first increasing and then
decreasing the bias.\cite{zas,gold,sollner,eaves,brown1,brown2,eaves1} This
effect is currently understood as caused by the interaction of the current
flowing through the device with the charge trapped in the well formed
between the barriers.\cite{gold} Calling Q this charge, its effect on the
incoming electrons may be viewed as an increase in the local potential by
Q/C, where C is the capacitance, lifting the resonance level in the same
amount. The latter can thus be somewhat above or below the emitter
conduction band edge at one same bias, depending on whether the well is
charged or uncharged. Since in the first case there is current flow through
the well and in the second there is none, both cases are physically
consistent, and in fact, observed experimentally. As an application, it has
been suggested that in the bistable region the device may act as a THz
detector and fast switch.\cite{ore,ore1}

Another instance are the THz oscillations that may arise in the presence of
a magnetic field in the direction of the current flow, which become chaotic
if the field is sufficiently strong.\cite{orellana,orella,brown3} In
contrast with the hysteresis effect described above, these oscillations are
associated with passage of the resonance across the emitter Fermi level.
Assuming the resonance is initially above the emitter Fermi sea and off
resonance, it will eventually enter the latter as the bias is increased,
allowing electrons to tunnel into the well. Thus, with a certain time
constant, charge begins to build up between the barriers. The Coulomb
repulsion between this charge layer and the incoming electrons effectively
cause an upward shift of the bottom of the well, and hence, also of the
resonance state. This may drive the system off-resonance again until the
charge, due to the finiteness of the barrier and the associated lifetime of
the resonance state, has tunneled out again into unoccupied states on both
the emitter and collector sides. The resonance then is no longer sustained
above the Fermi energy by the decreased charge in the well, the resonance
falls again and a new cycle begins. Whether this oscillation is damped away
or not is determined by the strength of the interaction, which in turn
depends on the projected density of states in the emitter at or near the Fermi
energy. By changing the profile in the density of states the magnetic field
enhances this coupling, and, if large enough, induces the system to perform
sustained oscillations. It is the purpose of this work to show that, under
favorable conditions, the oscillations may also be present in the absence of
a magnetic field, as well as situations of lower dimensionality such as
tunneling through a quantum wire or dot.

We model the system by a simple circuit that captures the essence of the
system under discussion, shown in Fig. 1. As will follow from the discussion
below, it includes the following features that appear to be essential in the
description of the device: (1) the existence of a resonance state with an
associated lifetime, which results in a time delay between the bias voltage
and the build-up or decay of the charge in the well (''tunneling time'');
(2) the existence of a Fermi sea on the emitter side, with a density of
states depending on the dimensionality of the device (bulk, planar or
linear), and on the magnetic field, if present; (3) the existence of the
Coulomb repulsion between the charge trapped in the well and the incoming
electrons. The nonlinear element in the circuit represents transmission from
the emitter side into the well, allowing a charge

\begin{equation}
\Delta Q(t,t+\tau)= \tau I_{max}f(V)
\end{equation}

\noindent to leave the source $U_0$ in the time interval $\tau$. The
dimensionless nonlinear function $f(V)$, depending on the voltage $V=U_0 - U$
between emitter and well, is the convolution of the transmittivity of the
resonance with the number of occupied states in the emitter, available for
tunneling. This latter quantity may be written as

\begin{equation}
N(V)=\theta(eV - E_r + E_F)\left[1 - \theta(eV - E_r)\right]\rho_d(V),
\end{equation}

\noindent where $\theta(x)$ is the Heaviside step function, $E_r$ is the
energy of the resonance level at zero bias and $E_F$, the Fermi energy. All
energies and voltages are assumed to be in electron-volts and measured with
respect to the bottom of the conduction band on the emitter side. The
quantity $\rho_d(V)$ represents the number of states in the emitter side,
with the component of kinetic energy along the current flow equal the
resonance energy shifted by the bias V. It is given by

\begin{equation}
\rho_d(V)=\alpha_d (eV+E_F-E_r)^{\frac{d-1}{2}}.
\end{equation}

\noindent Here $\alpha_d$ is a constant depending on the dimensionality $%
d=3,2$ or $1$ of the emitter (the well has dimension $d-1$). The values are, 
$\alpha_1=2, \alpha_2=2L\sqrt{2m^*}/\Pi\hbar, \alpha_3=m^*L^2/\pi\hbar^2$,
where $L$ is the width of the emitter and $m^*$ the effective mass of the
carriers.

The capacitance C represents the Coulomb barrier. As charge flows through
the nonlinear element, entering the well, the voltage drop $V=U_0-U$ is
reduced. At the same time charge is allowed to leave the well through the
collector represented by the load resistor R. These elements define a time
constant $\tau_0=RC$ characterizing the rate at which the well may be
emptied. With these definitions the conservation of charge provides an
equation of motion for the voltage $U(t)$ on the output side of the device,

\begin{equation}
C\left(U(t+\tau)-U(t)\right)=\Delta Q - \tau \frac{U(t)}{R}.
\end{equation}

\noindent Defining $V_n=V(t+n\tau)$ one obtains from Eqs. (1) and (4) the
following iterated map,

\begin{equation}
V_{n+1}=(1- \gamma) V_n - \gamma R I_{max}f(V_n) + \gamma U_0,
\end{equation}

\noindent where $\gamma=\tau/\tau_0$ and we assume that $\tau_0>\tau$ holds.
The map has fixed points given by

\begin{equation}
V^*=U_0-RI_{max}f(V^*).
\end{equation}

\noindent A simple linear analysis in the neighborhood of one of these fixed
points shows that for it to be stable the condition

\begin{equation}
-1< R I_{max} \frac{df}{dV}|_{V^*}<2/\gamma - 1
\end{equation}

\noindent must be satisfied. It follows that for either an increasing or
decreasing function $f(V)$ at the fixed point, an instability occurs
provided the local derivative is sufficiently large in absolute value for
the above condition to be violated. For a sharp resonance the function $f(V)$
will essentially follow the profile given by Eq.(3) within the window $E_r -
E_F\le V \le E_r$, becoming negligibly small elsewhere. The finite resonance
width smoothens the edges defined by the top and bottom of the Fermi sea.
Assuming a lorentzian form for the transmissivity $T(V)=\Gamma^2/[(E -E_r
+eV)^2 + \Gamma^2 ]$ one has for $\Gamma << E_F$,\cite{note}

\begin{equation}
f(V)\approx \rho_d(V) \left(arctg(\frac{eV - E_r + E_F}{\Gamma}) - arctg(%
\frac{eV - E_r}{\Gamma})\right).
\end{equation}

\noindent The expression in parenthesis is a positive definite funtion of V
resembling a hat, giving $f(V)$ such shape for d=1 since then $\rho_d(V)$ is
a constant; for d=2 and 3 $f(V)$ rather resembles an asymmetric hat. For
convenience we call it the "hat function" in what follows.

The fixed points, as determined by Eq. (6), may be found graphically as the
intersection of the straight line through the origin $V^*$, and the curve
representing the right hand side. Because of the inverted hat form of the
latter one can easily verify that there are either one (for large or small
value of $U_0$), or three, fixed points (some intermediate region). In this
last case the fixed point in the middle intersects the rather steep fall of
the $arctg$ function, yielding a likely unstable fixed point. The
intersection furthest to the right (RFP) occurs in the flat portion of the
hat function and is therefore stable.

We are interested in the leftmost fixed point (LFP) of the triple solution
case, representing the intersection located in the left side of the hat
function, which may also be present when there is just one fixed point. As $%
U_0$ varies, the intersection closely follows the functional form of the
number of states available for tunneling (3). We first consider the $d=1$
case, for which $\rho_1(V)$ is a constant in the relevant region. Then the
LFP traces the contour of the $arctg$ function and again the fall is
governed by the rather steep $arctg$, and, as with the RFP, will generally
violate the stability condition (7). The situation is as in the d=3 case
with magnetic field since the latter modifies the spectra in the emitter
transforming it into a series of quasi one dimensional dispersion laws, and
for which the instability is known to exist. \cite{orellana} For $d=2$,
there is a square root dependence at the intersect, and if $\Gamma$ is
sufficiently small the diverging local derivative will cause the stability
condition again to be violated near the edge (the Fermi energy).

The $d=3$ case needs closer attention because of the linear form of $%
\rho_3(V)$. Calculating the total emitter current flowing through the
resonance and relating the time constant $\tau$ with the resonance width
through $\tau \sim 2\pi\hbar/\Gamma$ one gets, away from the hat edges

\begin{equation}
R I_{max} \frac{df}{dV}|_{V^*}\sim \frac{e^2m^*D_z}{\epsilon\epsilon_0
\hbar^2\Gamma},
\end{equation}

\noindent where $D_z$ is the separation between the barriers, and $\epsilon,
\epsilon_0$ are the dielectric constants of the enclosed material and the
vacuum, respectively. Using the values appropriate for GaAs ($m^*=0.068m,
\epsilon=12.5$) one gets the simple stability condition

\begin{equation}
1.3 D_z(nm) < 2 - \gamma,
\end{equation}

\noindent where the distance $D_z$ is to be given in nanometers. This
condition is easily violated in actual samples.

We have done numerical studies in order to verify the presence of the
instability, and have found it to occur in all three cases discussed above.
Figure 2 shows the I-V curves for the (a) d=1, (b) d=2 and (c) d=3 cases in
the absence of a magnetic field. The parameters used in the figure are $%
\gamma = 0.5$, $\Gamma = 1 meV$, $E_F = 20 meV$, $E_r = 43 meV$ and $D_z =
13nm$, appropriate for typical devices based on AlGaAs. It is also necessary
to specify some short cross sectional dimensions in cases (a) and (b),
chosen as $D_xD_y = 100nm^2$ and $D_x = 10nm$ in the figure, respectively.
For these values, at low bias $U_0$ the device is always stable. Yet, as the
voltage increases and current starts to flow, marking the entrance of the
resonance in the Fermi sea, the iteration does not settle to a fixed point
but rather oscillates in steps $\tau$, first regularly (bifurcations region)
and then, at larger bias, in a chaotic fashion. The case with magnetic field
and $d=3$, not shown, produces a similar figure for large enough field, as
expected from results reported previously.\cite{orellana,orella}

In summary, we have shown that an instability induced by the interaction of
the current with the trapped charge may appear in resonant tunneling through
constrictions in one, two and three dimensions, when the resonance is close
to the Fermi level. Instead of integrating the quantum mechanical equations
of motion as done previously for the d=3 case with magnetic field \cite
{orellana}, we have used a circuit model with the advantage of mathematical
simplicity and the possibility of an analytical discussion of the stability
in one, two and three dimensions. Experimental accesibility of the
instability may require testing emission or absorption of THz radiation.
Usual electronics will normally just register a time average of the
oscillations because of the small value of their typical period, which, for
a resonance of 1meV width in a GaAs quantum well would be of the order of
4ps.

This research was supported in part by a C\'atedra Presidencial en Ciencia
and FONDECYT grant 1990425.

\begin{figure}[tbp]
\caption{Model circuit for a double barrier device. The function $f(V)$
represents the response for tunneling into the well}
\end{figure}

\begin{figure}[tbp]
\caption{Current-voltage iterates for a quantum well embedded in a (a) one
dimensional, (b) two dimensional and (c) three dimensional conductor}
\end{figure}

\end{document}